\documentclass[pra,showpacs%,preprint
]{revtex4}
\usepackage{amsmath}
\begin{document}
\title{Spin and Statistics in Nonrelativistic Quantum Mechanics:
      The Spin-Zero Case}

%%%%%%%%%%%%%%%\author{Murray Peshkin\\
%%%%%%%%%%%Physics Division, Argonne National Laboratory\\
%%%%%%%%%%%%%%%%Argonne, IL 60439 \\
%%%%%%%%%%%%%%%%%%%%%e-mail: peshkin@anl.gov}
\author{Murray Peshkin}
\affiliation{Physics Division, Argonne National Laboratory,
   Argonne, IL 60439}
\email{peshkin@anl.gov}

%%%%%%%%%%%%%%%%%%%%%%%%%%%%%%%%%%%%%%%%%%%%%%%%%%%%%%%%%%%%

\date{\today}

%%%%%%%%%%%%%%%%%%%%%%%%%%%%%%\maketitle
\begin{abstract}
It is proved from stated assumptions of nonrelativistic quantum
mechanics based on the Schr\"odinger equation that identical spin-zero
particles must obey symmetric statistics.
\end{abstract}

%%%%%%%%%%%%%%%%%%%%%%%%%%%%%%%%%%%%%%%%%%%%%%%%%%%%%%%%%%%%
\maketitle
%%%%%%%%%%%%%%%%%%%%%%%%%%%%%%%%%%%%%%%%%%%%%%%%%%%%%%%%%%%%

\section{INTRODUCTION}

The connection between spin and statistics (CSS) appears in
nonrelativistic quantum mechanics as a constraint on the solutions
of the Schr\"odinger equation. Practitioners usually introduce
coordinate and spin variables of individual particles as if the
identities of individual particles were observable. For integer
spin they then constrain the wave function to be symmetric under
exchange of the labels of any pair. For half-integer spin, the
wave function is made antisymmetric.  Equivalently, the total spin
$S$ and the relative orbital angular momentum $\ell$ of each pair
are constrained to be both even or both odd. Alternatively, the
machinery of Fock space is introduced with creation and
annihilation operators obeying commutation relations designed to
achieve the same end~\cite{L}. To date, the available experimental
information confirms the CSS but one may still ask whether quantum
mechanics requires that all particles must obey it.

Pauli's \cite{P} original proof of the CSS depended upon the somewhat
hazardous assumptions of local relativistic quantum field theory.  That
was followed by similar proofs based on more modern versions of quantum
field theory \cite{W}.  Feynman, in his lectures \cite{F}, expressed the
dissatisfaction with that kind of proof by saying, ``$\ldots$ An
explanation has been worked out by Pauli from complicated arguments of
quantum field theory and relativity $\ldots$ we have not been able to
find a way of reproducing his arguments on an elementary level.  This
probably means we do not have a complete understanding of the
fundamental principle involved."  Duck and Sudarshan \cite{DS}, in their
recent extensive critical review of the relevant theoretical and
experimental literature, found that Feynman's challenge had not been
answered satisfactorily.

Messiah and Greenberg \cite{MG} investigated the status of the CSS
within nonrelativistic quantum mechanics by rigorously analyzing
the consequences of the particles' being identical for the
behavior of wave functions under permutations.  They found that
the minimal assumptions of quantum mechanics that they used permit
symmetric, antisymmetric, and intermediate statistics for
identical particles of all spins. 

Leinaas and Myrheim \cite{LM} introduced a new approach by taking
seriously the assumption that the quantum mechanical variables
representing physical observables should stand in one-to-one
correspondence with those observables.  For two identical spinless
particles with no other observables the configuration space consists of
the unordered pairs of vectors $\{ \mathbf{r}_1, \mathbf{r}_2 \}$, for
which the subscripts label points in space, not particles, and $\{
\mathbf{r}_2, \mathbf{r}_1 \}$ represents the same point in the
six-dimensional configuration space as does $\{ \mathbf{r}_1,
\mathbf{r}_2 \}$.  For particles with non-zero spins, the spin
variables must also be included, but that complication will not
necessary for present purposes. Leinaas and Myrheim were able, by
analyzing phases introduced by parallel transport in the so-defined
space of $\{ \mathbf{r}_1, \mathbf{r}_2 \}$, to eliminate intermediate
statistics, but not to choose between symmetric and antisymmetric
statistics for particles of any spin. Their work was later extended by
Berry and Robbins \cite{BR1,BR2}, who gave it a rigorous mathematical
foundation for all spins and showed exactly how different allowable
assumptions about parallel transport relate to geometrical properties of
the configuration space, including the geometrical Berry phase.  The
outcome for the connection between spin and statistics was, however, the
same. Particles of any spin can have symmetric or antisymmetric
statistics; neither is excluded by the physical assumptions that they
used.

This paper is limited to the case of spinless particles.  I follow
Leinaas and Myrheim by identifying $\{ \mathbf{r}_2, \mathbf{r}_1 \}$
with $\{ \mathbf{r}_1, \mathbf{r}_2 \}$, but I additionally assume that
the wave function $\Psi(\{\mathbf{r}_1, \mathbf{r}_2\})$ must be a
continuous function of $\mathbf{r}_1$ and $\mathbf{r}_2$ because of the
second derivatives in the Schr\"odinger equation. Those assumptions lead
unambiguously to the result that identical spinless particles must be
bosons, not fermions. The method used here is elementary, involving only
the properties of rotation and angular momentum. Parallel transport and
its connection with global geometric properties of the configuration
space are not needed.

\section{ASSUMPTIONS}

\emph{Assumption 1.}  The only independent dynamical variables are the
positions and momenta of the particles.  There are no spins or other
internal variables.  Then the wave function for a single particle has
one component and is scalar under rotation.  In other words,
$\Re\Psi(\mathbf{r})=\Psi(\Re^{-1}\mathbf{r})$ for any rotation $\Re$.

\emph{Assumption 2.}  For any two spinless particles, the wave
functions are products of scalar, one-component single-particle
wave functions or linear combinations of such products.  For
identical spinless particles, the configuration space is that of
Leinaas and Myrheim, consisting of the unordered pairs $\{ \mathbf{r}_1,
\mathbf{r}_2 \}$, and the wave functions are scalar, one-component
functions $\Psi(\{\mathbf{r}_1, \mathbf{r}_2\})$, where the
subscripts label points in space, not particles.  This assumption
is, as mentioned above, motivated by the principle of quantum
mechanics that the dynamical variables should be observable, at
least in principle.

\emph{Assumption 3.}  The wave function $\Psi(\{\mathbf{r}_1,
\mathbf{r}_2\})$ is a continuous function of the variables
$\mathbf{r}_1$ and $\mathbf{r}_2$. Continuity is required by the
appearance of $\nabla_1^2$ and $\nabla_2^2$ in the Hamiltonian. It is in
fact possible to avoid continuity by making a gauge transformation that
results in a discontinuous vector potential in the Hamiltonian, but that
will not change the physics and I will not consider it further.

\emph{Assumption 4.} In extending the conclusions from two
identical particles to many, I will make use of an assumption of
asymptotic separability to be explained below.

\section{Two Spinless Particles}

The domain of the relative coordinate $\mathbf{r} = \mathbf{r}_{1} -
\mathbf{r}_{2}$ is only half of three-dimensional space because
$\mathbf{r}$ and $-\mathbf{r}$ cannot both be in that domain since no
observation can distinguish between them. For fixed $r=|\mathbf{r}|$,
Leinaas and Myrheim represent that domain graphically by a hemisphere of
radius $r$ whose base is a circle in the $z=0$ plane but with only half
of that circle included.
\begin{equation}
   \Psi(\{\mathbf{r}_1, \mathbf{r}_2\})
   = \sum _{\ell m} \, a_{\ell m}(\mathbf{R},r) \,
                        Y_{\ell m} (\vartheta ,\varphi )
   \>,
\end{equation}
where $\mathbf{R}$ is the center-of-mass coordinate and the domain
of the relative coordinate $\mathbf{r}$ is given by
\begin{gather}
   r \ge 0 \>,
   \\ \notag
   0 \le \vartheta \le \frac{\pi}{2} \>,
   \\ \notag
   0 \le \varphi < \left \{
   \begin{matrix}
      2 \pi, & \mathrm{for} \ \vartheta < \pi /2 \>, \\
      \pi, & \mathrm{for} \ \vartheta = \pi /2 \>.
   \end{matrix}
   \right .
\end{gather}

The domain of $\mathbf{r}$ defined here differs technically from that of
Leinaas and Myrheim in that the one defined here is simply connected.
No two points $\mathbf{r}$ and -$\mathbf{r}$ need to be identified in
that domain because -$\mathbf{r}$ is never in the domain of
$\mathbf{r}$.  The choice of the hemisphere based on the $z$ axis is of
course arbitrary.  Any other hemisphere would represent the same
physics.

The sets $Y_{\ell m}$ for even and for odd $\ell$ are separately
complete on the hemisphere. It is proved in Appendix \ref{app:A}
below that even values and odd are superselected from each other;
both cannot appear in one wave function.  What remains is to
eliminate one of them.

Consider a point $\{ \mathbf{r}_1, \mathbf{r}_2 \}$ corresponding
to
\begin{equation}
   \mathbf{r}_1=(x,y,\varepsilon)
   \>,
   \qquad
   \mathbf{r}_2=(-x,-y,-\varepsilon)
   \>,
   \qquad
   \mathbf{r}=(2x,2y,2\varepsilon)
   \>.
\end{equation}
For infinitesimal $\varepsilon$, that point must be
infinitesimally close to the point
\begin{equation}
   \mathbf{r}_1=(x,y,-\varepsilon)
   \>,
   \qquad
   \mathbf{r}_2=(-x,-y,\varepsilon)
   \>,
   \qquad
   \mathbf{r}=(-2x,-2y,2\varepsilon)
   \>.
\end{equation}
The two points have the same $r$ and $\vartheta$,
but differ in $\varphi$ by $\pi$. Continuity under changes in
$\mathbf{r}_1$ and $\mathbf{r}_2$ requires
\begin{equation}
   \Psi(\mathbf{R}, r,\vartheta,\varphi) \ \rightarrow \
   \Psi(\mathbf{R}, r,\vartheta,\varphi \pm \pi)
   \>
\label{eq:3}
\end{equation}
as $\vartheta \rightarrow \pi/2$ from below.
\begin{gather}
   \Psi(\mathbf{R}, r,\vartheta,\varphi) \ \rightarrow \
   \sum _{\ell m} \, a_{\ell m}(\mathbf{R},r) \,
                        Y_{\ell m} \left ( \frac{\pi}{2} ,\varphi \right )
   \>
\label{eq:4}
   \\ \notag
   \Psi(\mathbf{R}, r,\vartheta,\varphi \pm \pi) \ \rightarrow \
   \sum _{\ell m} \, (-1)^m \, a_{\ell m}(\mathbf{R},r) \,
                        Y_{\ell m} \left ( \frac{\pi}{2} ,\varphi \right )
   \>
\end{gather}
Eqs.(\ref{eq:3},\ref{eq:4}) are consistent only if  $Y_{\ell m}
\left ( \frac{\pi}{2} ,\varphi \right )$ vanishes for all odd
values of $m$. That is the case when $\ell$ is even but not when
$\ell$ is odd. Therefore, $\ell$ must be even and the two
identical spinless particles must have symmetric statistics.

\section{Discussion}

It has been proved, under stated general assumptions of quantum
mechanics, that two identical spinless particles with no internal
degrees of freedom must have even relative orbital angular momentum,
which implies that they are bosons, not fermions. This proof did not
make use of relativity or of quantum field theory.  The approach used
here is based on the requirement that the point $\{ \mathbf{r}_1,
\mathbf{r}_2 \}$ in the configuration space for two identical spinless
particles is the same point as $\{ \mathbf{r}_2, \mathbf{r}_1 \}$.  This
approach was enabled to go beyond previous work departing from the same
requirement and to find unambiguously that the two spinless particles
are bosons by the introduction of the additional requirement that wave
functions must be continuous under variations of the particle
coordinates $\mathbf{r}_1$ and $\mathbf{r}_2$ because of the second
derivative in the Hamiltonian and the Schr\"odinger equation, and by
consideration of the relative orbital angular momentum.  The continuity
in $\mathbf{r}_1$ and $\mathbf{r}_2$ was used to relate the wave
function at a relative coordiate $\mathbf{r}$ near the relative $z$=0
plane to the wave function at a rotated point.  No question of
multiple-valued wave functions arose because the domain of $\mathbf{r}$
is simply connected.

The extension of the spin-statistics connection to many particles has
been given in a general way by Berry and Robbins \cite{BR1}.  Here, to
complete the discussion of spinless particles, I give a simple heuristic
proof that should apply to any theory that is asymptotically separable
in the sense that moving all particles except two to a great distance is
the same as removing them; the motion of the two remaining particles is
unaffected by the presence or absence of other particles a great
distance.  For many particles, the configuration space consists of the
unordered multiplets $\{ \mathbf{r}_1$, $\mathbf{r}_2$, $\mathbf{r}_3$,
$\ldots$, $\mathbf{r}_N \}$, where the subscripts label points in space,
not particles. Select any pair of $\mathbf{r}_j$ and consider a wave
function $\Psi(\{\mathbf{u}, \mathbf{v}, s \mathbf{w} \} )$, where
$\mathbf{u}$ and $\mathbf{v}$ are the two selected $\mathbf{r}_j$,
$\mathbf{w}$ stands for all the other $\mathbf{r}_j$, and $s$ is a scale
factor. For sufficiently large $s$, it is assumed that the dynamics in
the neighborhood of $\mathbf{u}$ and $\mathbf{v}$ is unaffected by the
existence of the remaining particles. Then the relative orbital angular
momentum $\ell$ of the particles at those points must be even. Now
reduce $s$ continuously to $s=1$.  If the wave function is to be
continuous, $\ell$ values cannot jump so they must remain even.  Then
the relative angular momentum of each pair is even and the particles are
bosons.

This paper addresses the motion of identical particles in all of
three-dimensional space.  Boundary conditions, even ones that confine
two particles to two separate boxes, appear not to challenge the results
because such boundary conditions are idealizations.  In physical reality
they can be replaced by sufficiently high potential barriers, whose
existence is not precluded by the the assumptions used here.  The same
is true of the Aharonov-Bohm effect \cite{PT}, where real flux lines
have nonvanishing thickness and nonvanishing penetrability.  The
hypothetical case of the motion of charged particles near a Dirac
monopole, where the multiply-connected domain of the individual particle
coordinates is not merely an idealized limit, is not covered by the
results found here, nor apparently in other treatments of the connection
between spin and statistics.

The methods used here are not directly applicable in their present
simple form to particles with spin because the spatial continuity
condition alone is insufficient to determine the relative phase of
$\Psi(\varphi)$ and $\Psi(\varphi \pm \pi)$ at $\vartheta= (\pi/2) -
\varepsilon$ when $\Psi$ contains spinors in addition to its spatial
variables.

Berry and Robbins, in what they call a perverse case, give a wave
function for two identical spinless fermions which in the
present notation would be
\begin{equation}
   \Psi(\mathbf{R}, r,\vartheta,\varphi) \ = \
   \frac{\mathbf{r}}{r} \
   \sum _{\ell m} \, a_{\ell m}(\mathbf{R},r) \,
                        Y_{\ell m} \left ( \frac{\pi}{2} ,\varphi \right )
   \>.
\end{equation}
Under the standard assumptions of quantum mechanics used here, this is
not an admissible wave function for two spinless particles because it is
a three-component vector, not a one-component scalar, under rotation.

The proof of the CSS in nonrelativistic theory differs fundamentally
from the proofs in quantum field theory.  In nonrelativistic theory,
states with different numbers of identical particles are superselected
from each other and live in spaces with different topologies.  In
quantum field theory states with different particle numbers are not
superselected and the spatial variables are only labels and have the
same topology for all states.

\section*{ACKNOWLEDGMENTS}

I thank Michael Berry for valuable communications and criticisms.  This
work is supported by the US Department of Energy, Nuclear Physics
Division, under contact No. W-31-109-ENG-38.

\appendix

\section{Superselection of even from odd angular momentum}
\label{app:A}

Superselection of even from odd $\ell$ is required by the
condition that the rotation of two spinless particles through
angle $\pi$ about an axis through their center of mass and
perpendicular to their relative coordinate must restore their
initial physical state. Consider an initial state $| \mathbf{r}_0
\rangle$ with $0 < \vartheta_0 < \pi/2$.  For this purpose, the
irrelevant center-of- mass coordinate $\mathbf{R}$ is suppressed.

Define unit vectors $\mathbf{n}_j$ and ``body-fixed" angular
momentum projections $K_j$  by
\begin{gather}
    \mathbf{n}_3 = \frac{\mathbf{r}_0}{r_0}
    \>,
    \qquad
    \mathbf{n}_1 = \frac{\mathbf{\hat{z}} \times \mathbf{r}_0}
                        {|\mathbf{\hat{z}} \times \mathbf{r}_0 |}
    \>,
    \qquad
    \mathbf{n}_{2} = -\mathbf{n}_{3} \times \mathbf{n}_{1}
    \>,
    \notag \\
    K_{j} = \mathbf{n}_{j} \cdot \mathbf{L}
    \>,
    \\
    [K_{i} ,K_{j} ] = i\varepsilon _{ijk} K_{k} \>,
    \notag \\
    K^{2}_{1} + K^{2}_{2} + K^{2}_{3} = \mathbf{L}^2
    \>,
\label{eq:A2}
    \\
    | \mathbf{r} _{0} \rangle =
    \sum _{\ell \mu} \ \alpha _{\ell \mu}(r_0) \
                       | \ell \mu \rangle_1
    \>,
\label{eq:A3}
    \\ \notag
    K_1 \ | \ell, \mu \rangle_1 = \mu \ | \ell, \mu \rangle_1
    \>,
    \qquad
    \mathbf{L}^2 \ | \ell, \mu \rangle_1 =
    \ell (\ell + 1) \ | \ell, \mu \rangle_1
    \>.
\end{gather}
$K_{1}$ generates the rotations of $| \mathbf{r}_{0} \rangle$
about an axis perpendicular to $| \mathbf{r}_{0} \rangle$. A
rotation of $\mathbf{r}_1$ and $\mathbf{r}_2$ through angle $\pi$
around that axis must restore the initial state except for a
possible phase factor $e^{\mathrm{i} \delta}$. In terms of the
relative $\mathbf{r}$, that rotation appears as a rotation through
angle $\left ( \frac{\pi}{2} - \vartheta_0 \right )$ down to the
$z=0$ plane, followed by a rotation through $\left ( \frac{\pi}{2}
+ \vartheta_0 \right )$ for a total angle of $\pi$ to restore $|
\mathbf{r}_{0} \rangle$
\begin{equation}
   e^{\mathrm{i} \pi K_1} \ | \mathbf{r}_0 \rangle =
   \sum _{\ell \mu} \ \alpha_{\ell \mu}(r_0) \
   e^{\mathrm{i} \pi \mu} \ | \ell \mu \rangle
   \ = \
   e^{\mathrm{i} \delta} \ \sum _{\ell \mu}
   \alpha_{\ell \mu}(r_0) \ | \ell \mu \rangle
   \>.
\label{eq:A4}
\end{equation}
In writing Eq.(\ref{eq:A4}), I have implicitly assumed that the angular
momentum eigenfunctions are continuous under infinitesimal changes of
$\mathbf{r}_1$ and $\mathbf{r}_2$, as in Section III above.

Eqs.(\ref{eq:A2}) constrain the values of $\mu$ to be integers or
integers plus one-half. Then, Eq.(\ref{eq:A4}) requires that
\begin{equation}
   \mu = 2n + \delta
   \>,
\end{equation}
where the values of $n$ are integers and $\delta$ is zero, one-half,
one, or one-and-one-half. If states with different $\delta$ exist, they
are superselected from each other and cannot appear in the same wave
function.  In other words, even $\mu$ are superselected from odd.

The state $| \mathbf{r}_0 \rangle$ is an eigenfunction of $K_3$
with eigenvalue equal to zero.
From Eqs.(\ref{eq:A2}),
\begin{equation}
   | \mathbf{r}_0 \rangle \ = \
   \sum_\ell \ \beta_\ell \ | \ell, 0 \rangle_3
   \>,
\label{eq:A7}
\end{equation}
for some $\beta_\ell$, where $| \ell, 0 \rangle_3$ is an
eigenfunction of $\mathbf{L}^2$ and $K_3$ with eigenvalues
$\ell(\ell+1)$ and zero, respectively. Then the values of $\ell$,
and consequently of $\mu$, must be integers.

The eigenfunctions of $K_1$ are related to those of $K_3$ by a
rotation through angle $(\pi/2)$ around the $\mathbf{n}_2$ axis.
Therefore, from Eq.(\ref{eq:A7}),
\begin{equation}
   | \mathbf{r}_0 \rangle \ = \
   \sum_{\ell \mu} \ D^{\ell}_{\mu 0}(\Re) \ \beta_\ell \ | \ell, \mu \rangle_1
   \>.
\label{eq:A8}
\end{equation}
Here, $\Re$ is a rotation through angle $(\pi/2)$ around the
$\mathbf{n}_2$ axis, which carries $\mathbf{n}_3$ into $\mathbf{n}_1$,
and $D^{\ell}(\Re)$ is the rotation matrix for wave functions of angular
momentum $\ell$.  For the $\pi/2$ rotation, all the $\ D^{\ell}_{\mu 0}(\Re)$
vanish except for even values of $\ell$-$\mu$. Then, since even $\mu$
are superselected from odd, even $\ell$ must likewise be superselected
from odd.

This proof has been given for a single $| \mathbf{r}_0 \rangle$,
but continuity assures that one selection of even versus odd
applies to all $| \mathbf{r}_0 \rangle$.

\end{document}